\documentstyle[preprint,aps]{revtex}
\begin{document}
\draft
\tightenlines

\title{An Elastic Lattice in a Random Potential}
\author{Eugene M. Chudnovsky and Ronald Dickman$^{\dag} $}
\address{Department of Physics and Astronomy, Lehman College, CUNY,
Bedford Park Boulevard West, Bronx, NY 10468-1589}
\date{\today}
\maketitle
\begin{abstract}
Using Monte Carlo simulations, we study the properties of an elastic 
triangular lattice subject to a random background potential. 
As the cooling rate is reduced, we observe a
rather sudden crossover between two different  
glass phases, one with exponential decay of translational correlations,
the other with power-law decay. Contrary to predictions
derived from continuum models, no 
evidence of a crossover in the mean-square displacement, $B(r)$, 
from quadratic growth at small $r$,  to logarithmic growth at 
large $r$ is found.
\vspace {0.3truecm}

\end{abstract}
\vspace{1.0truecm}

$^{\dag}$e-mail address: dickman@lcvax.lehman.cuny.edu

\newpage

The structure of  an elastic lattice in a random background has received
much attention in recent years, due to its relevance to such systems as a
Wigner crystal in a 
semiconductor with impurities  \cite{ANDREI}, a charge density wave in a 
weakly disordered medium \cite{DAI}, an atomic monolayer on an imperfect 
crystal surface \cite{NAGLER}, a magnetic bubble lattice in a ferromagnetic 
film with defects \cite{SESHARDI}, and a vortex lattice in a disordered 
superconductor  \cite{MURRAY}. The latter problem has been  investigated
most intensively, both theoretically and experimentally, because translational 
correlations in the vortex lattice are responsible for such important properties 
of superconductors as resistivity and critical current.  

Most theoretical works on elastic lattices in a random background have
concentrated on a continuum approach in which the deformation of the 
lattice is described by the displacement field ${\bf u}({\bf r})$. In this 
model the background is described by a random potential  
$V[{\bf r}, {\bf u}({\bf r})]$ that satisfies  
$V[{\bf r}, {\bf u}({\bf r})+{\bf a}_i ]=V[{\bf r}, {\bf u}({\bf r})]$
which accounts for the periodicity of the lattice.  (${\bf a}_{i}$ is a 
lattice vector.) The energy of this system is 
\begin{equation}
U=\int\,d^{d}r\,[{\hat{\alpha}}{\nabla}{\bf u}{\nabla}{\bf u}+
V({\bf r}, {\bf u})]\;\;\;,
\end{equation}
where the first term represents elasticity; ${\alpha}_{iklm}$ being the tensor
of elastic moduli. When $V=0$, that is in the absence of pinning, the 
energy is minimized by ${\bf u}= const.$ which represents perfect translational
order. In the presence of pinning the lattice develops deformations,
${\bf u}({\bf r})$, and the question arises whether (and how rapidly) these 
deformations destroy long range translational correlations. A convenient
measure of the disorder is
\begin{equation}
B({\bf r})=<[{\bf u}({\bf r})-{\bf u}(0)]^{2}>
\end{equation}

If pinning is weak and only a small area of the lattice is of interest,
the deformation $u$ is small compared to the
lattice spacing $a$. Then $V({\bf r}, {\bf u})$ can be written as
$-{\bf f}({\bf r}){\cdot}{\bf u}$ and the problem reduces to the random
force problem, for which the solution is known: a random force, 
no matter how weak, destroys the long range translational order in less than 
four dimensions.  
Simple statistical arguments yield \cite{LARKIN,IMRY}
$B{\sim}(f^{2}/{\alpha}^{2})r^{4-d}$, where $f^2$ denotes the variance of
the random force.  That is,
$B{\propto}r$ in three dimensions and $B{\propto}r^{2}$ in two
dimensions.  Equating $B$ to
$a$ yields an estimate of the translational correlation length,
$\xi \propto f^{2/(d-4)}$. This estimate seems to be quite robust with
respect to approximations and assumptions about the random potential. 
It has been pointed out, however, that the random field model cannot 
provide the correct rate at which translational correlations are destroyed
for $r>{\xi}$, as it does not take into account the periodicity of the lattice.
More sophisticated approaches  based upon the Gaussian variational 
method \cite{MEZARD} and the functional renormalization group \cite{FISHER}
suggest that $B \propto (4-d)\ln r$ for $r> \xi$ 
\cite{NATTERMANN,GIAMARCHI,BOUCHAUD}. Another
renormalization group approach \cite{CARRARO} suggests that the 
disordered lattice freezes below a certain temperature into a glass state with 
$B \propto \ln^{2} r $. The upshot of  these analytical results is  a much 
slower decay of translational correlations than predicted by the random field 
model \cite{LARKIN,IMRY}. 

The purpose of this paper is to compare the above predictions of the 
continuous  model with Monte Carlo simulations of a discrete triangular
lattice subject to a random background potential. As in the continuous
approaches mentioned above, we limit our consideration to
lattices free of dislocations. The results obtained under this assumption
may be relevant to real systems since patterns of vortex lattices observed
in decoration  experiments show remarkably large areas free of dislocations.  

We consider a two-dimensional triangular lattice of particles coupled by a 
harmonic, nearest-neighbor interaction, and subject to a static, random potential
$V(x,y)$.  Let ${\bf x}_{i} = (x_i,y_i)$ 
denote the position of particle $i$.
Then the potential energy of the system is 
\begin{equation}
E =\frac{1}{2} \sum_{<i,j>} (r_{i,j} - 1)^2 + \sum_i V({\bf x}_i) ,
\;\;\;,
\label{energy}
\end{equation}
where the first sum is over all nearest-neighbor pairs 
in the triangular lattice, and
$r_{i,j} = |{\bf x_i} - {\bf x_j}| $.
(All quantities are dimensionless in our 
formulation, with the basic length scale $a=1$ set by the nearest-neighbor (NN)
separation in the unstrained lattice and the basic energy scale $e=1/2$ set by the
energy of a NN pair with $|r - 1| = 1$, so that the spring constants have unit
magnitude.)  

To avoid severe distortions of the lattice we impose a
{\em planarity constraint}, which prevents any particle from escaping the ``cage"
defined by the current positions of its six nearest neighbors.  
(This is enforced by demanding that if {\bf x}, {\bf y} and {\bf z} form a
unit triangle in the unstrained lattice, then the angle between {\bf x}-{\bf y}
and {\bf z}-{\bf y} not exceed $\pi$.)
We adopted this constraint in preference to equipping the particles with
hard cores, since hard cores of a size sufficient to prevent this kind of distortion
yield a rather strong anharmonicity.  In the present model the anharmonicity is
weak, with the deviation from equipartition amounting to $< 5\%$ at the temperatures
of interest.

The background potential $V(x,y)$ is generated in two steps.  We first 
generate $R(i,j)$, an $N \times N$ array (we used $N=101$) of uncorrelated random
numbers, uniformly distributed on [-1/2, 1/2].  We smooth this array by replacing
each entry $R(i,j)$ by the sum of the entry and its four nearest-neighbors, using periodic
boundaries at the edges.  The smoothing process is applied a total of three times, so that
elements up to six units apart have a nonzero correlation.  (Each entry of $R$ is now a
weighted sum of 25 random numbers, and so represents a good approximation
to a Gaussian random variable.)  To find $V(x,y)$, we define a random, piece-wise
constant (on a scale of $\sim 10^{-3}$) map between points $(x,y)$ and a set
of four entries $R_1,...,R_4$, where $R_k \equiv R(i_k,j_k)$.  The first entry is
given by the modular expression: $i_1 = [ 7x \; (mod \;101) ]$, where the brackets 
denote the largest integer, and similarly for $j_1$.
Then $i_2 = [f i_1 \; (mod \;100)] $, where 
$f \equiv 2049 |R_1|$ is a random multiplier; $j_2$ is
defined similarly, yielding $R_2$.  We repeat the process, this time using
$f = 513 |R_1 + R_2|$, and then once more, with $f = 257 |R_1 + R_2 + R_3|$.
The resulting $V(x,y) \equiv A \sum_{i=1}^4 R_i$ is an (approximately) Gaussian
random field with short-range correlations; 
$C_V (r) \equiv \langle V({\bf x})V({\bf y}) \rangle_{|{\bf x} - {\bf y}| = r} /\sigma^2 $
drops from unity to about 0.3 for $r \simeq 10^{-3}$, and then decays in
roughly linear fashion, remaining essentially zero for $r \geq 0.6$.  
(We found that choosing prime factors in the multipliers hastened the 
decay of correlations.) The 
strength of the background potential is characterized by its standard
deviation, $\sigma$, which we control by varying the constant $A$.
($\sigma = 0.2$ and 0.5 are used in this study.  All simulations were performed
on DEC Alpha workstations, and employed the random number generator
supplied with the machine.)

Thus the background potential differs from
the kind typically employed in studies of off-lattice systems subject to quenched
randomness which distributes a certain density of identical centers of force,
with potential $v(r)$,
at random positions, {\bf q}$_i$, and sets $V({\bf x}) = \sum_i v(|{\bf x}-{\bf q}_i|) $.
Our potential is a closer approximation to the Gaussian random field
used in theoretical analyses.

We simulated hexagonal-shaped lattices of $M$ particles to a side
(a total of $3M(M-1) + 1$ particles),
with open boundaries.  We report results  for $M=60$ and
$M=120$.
We used open boundaries to eliminate global periodicity 
as a restraint on the growth of
particle displacements.  
In each step of the simulation, a particle is selected at random and subjected 
to a trial displacement uniform on a square of side $D=0.5$, symmetric about the
origin. The move is accepted if the total change in energy $\Delta E \leq 0$;
if  $\Delta E $ is positive the new position is accepted with
probability $e^{-\Delta E/T}$.  Our time unit comprises one attempted
move per particle.  A preliminary study of the lattice without the random
background potential (see also Ref. \cite{DICKMAN}) revealed that the 
correlation function
\begin{equation}
g_{\bf G}({\bf r})= \langle e^{i{\bf G}{\cdot}({\bf x}_i - {\bf x}_j ) } \rangle
\;\;\;,
\label{trcorr}
\end{equation}
(the thermal average is over all pairs with ${\bf x}_i - {\bf x}_j = {\bf r} $ in the
unstrained lattice; {\bf G} is a reciprocal lattice vector), shows a power-law decay,
$g_{\bf G}({\bf r}) \sim r^{-\eta} $ with $\eta$ proportional to temperature, $T$,
as expected \cite{HALPERIN}.

Our primary interest is in the behavior of the mean-square displacement from
equilibrium,
\begin{equation}
B({\bf r}) \equiv <[{\bf x}_i - {\bf x}_j - {\bf r}]^2>
\;\;\;,
\label{msqd}
\end{equation}
where we average over all pairs with ${\bf x}_i - {\bf x}_j = {\bf r} $ in the
unstrained lattice, and the angle brackets denote an average over disorder.
To avoid the effects of strong distortions that may appear at the 
boundary, we only average over particles
at least $M/2$ sites distant from it. Configurations 
are generated by taking the perfect, unstrained lattice and permitting it
to relax, in the presence of $V({\bf x})$, for $t_A = 10^4$ --- $10^5$ 
timesteps at temperature 1.
After this ``annealing" phase the system is gradually cooled: the
inverse temperature $\beta = 1/T$ increases at a constant 
cooling rate $\Gamma$, until $T=0.01$, at which time we
compute $B$ and other properties.  (At this point the system is essentially
at temperature zero, since the typical background energy $\sigma \gg T$.)

Depending on the cooling rate, we observe two qualitatively distinct kinds of
$B(r)$.  In Fig. 1, for example, we show $B(r)$ for $M=60$, 
$\sigma =0.2$, $t_A = 10^4$, and a variety of different cooling rates.
For relatively rapid cooling, ($\Gamma > 4 \times 10^{-4}$ for the parameters
of Fig. 1), $B$ grows $\sim \ln r$ for small $r$ before crossing
over to some slower growth whose precise form is unclear.  
The correlation length $\xi$, defined via $B(\xi) = 1$, decreases with $\Gamma$
and in many cases exceeds the system size $M$. The logarithmic growth is
reminiscent of a lattice in {\em thermal equilibrium} \cite{HALPERIN,DICKMAN},
suggesting that in
this case the system has undergone a kind of glass transition, the final
$B(r)$ being a remnant of the thermal disorder when the lattice fell out of
equilibrium with the heat bath.  In this situation the background potential
serves to ``freeze" thermal fluctuations, even though
the system does not have 
time to optimize its configuration with respect to $V({\bf r})$.  Fig. 2
illustrates the similarity between $B(r)$ in thermal equilibrium and in a
rapidly cooled system.  

For lower cooling rates, the particles have the opportunity to explore more
of the local potential energy landscape, and the lattice distortion, while
more modest for small $r$, becomes sizeable on large scales.  
In this regime the correlation length  {\em decreases} as we reduce $\Gamma$;
increased exploration results in greater distortion.  An important open
question concerns the nature of $B(r)$ in the ground state: Does it grow $\sim r^2$
indefinitely, or cross over to a slower, perhaps logarithmic growth law?
As we reduce the cooling rate, and hence probe nearer the ground state
of the model, we find that $B(r)$ maintains a faster than linear growth.
Fig. 3 shows that in this regime, $B({\bf r}) \sim r^2$ to a good approximation.
In no instance do we observe a crossover to logarithmic growth at large $r$, even 
though our data extend to $r/\xi \simeq 14 $ for some systems.

Fig. 4 shows the elastic and potential energies ($u$ and $v$, respectively),
per particle as a function of $\Gamma$, for the conditions of Fig. 1.  While
there are some fluctuations (due to computer-time restrictions we average over
sets of only five independent disorder configurations in this series of studies), 
the energy decreases
with cooling rate, as expected.  We also plot $B(r=50)$ as a measure of
the character of the final state (quenched thermal fluctuations versus
optimized in the random background).  It appears
that the change in the
behavior of $B$ occurs rather suddenly, at $\Gamma \simeq 4 \times 10^{-4}$.

In summary, we studied the mean-square displacement in an elastic lattice
subject to a static random potential, and find no evidence of crossover to 
logarithmic growth in the well-relaxed regime.  A crossover between two different glass
phases, one with exponential decay of translational correlations, the other
having power-law decay, is observed as the cooling rate is reduced.
 
This work was supported by the Department of Energy
under Grant No. DE-FG02-93ER45487.

\pagebreak

\noindent{\bf Figure Captions}
\vspace{2em}

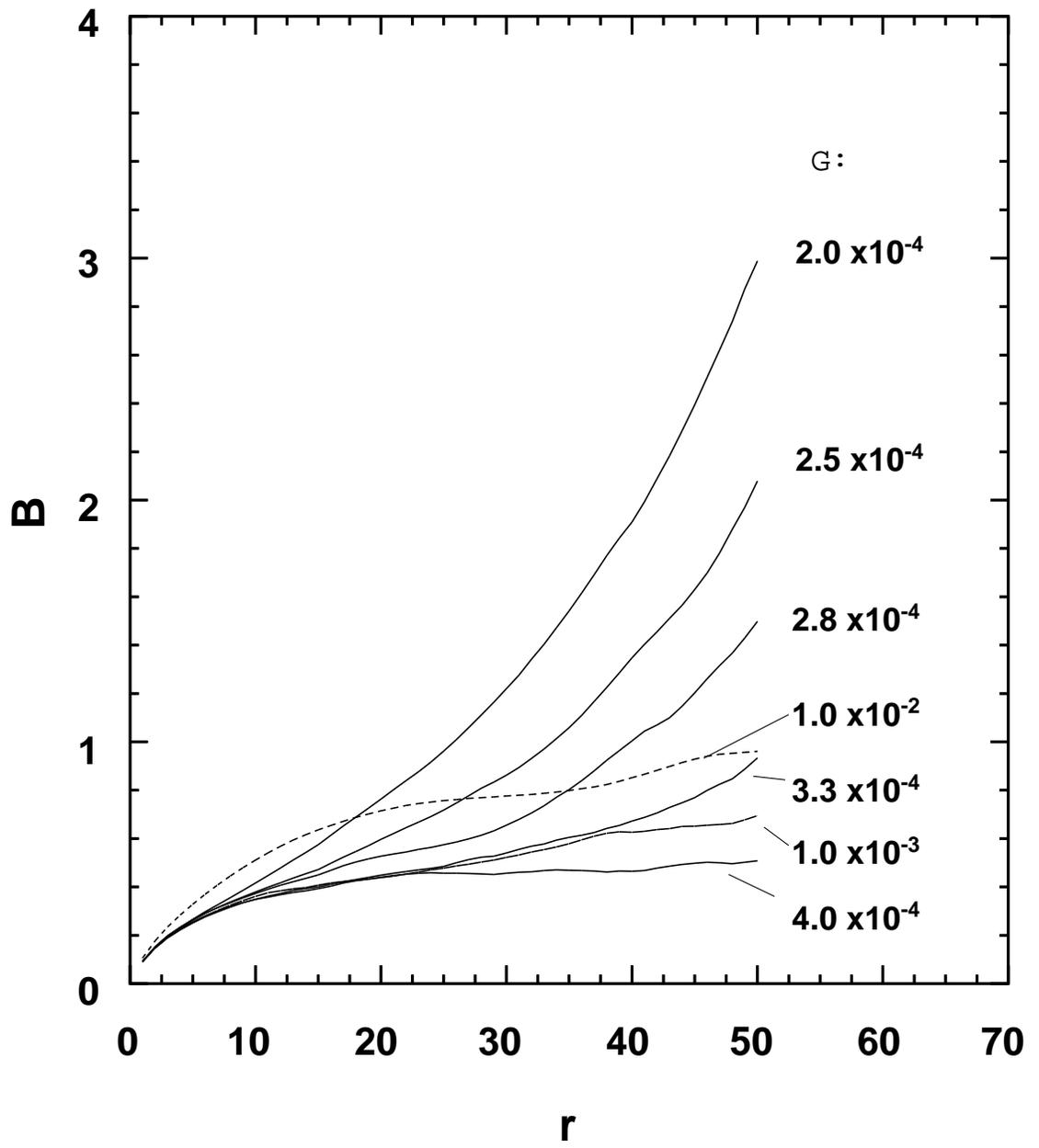
\noindent FIG. 1. $B(r)$ versus $r$ for $M=60$, 
$\sigma =0.2$, $t_A = 10^4$, and various cooling rates $\Gamma$.
\vspace{1em}

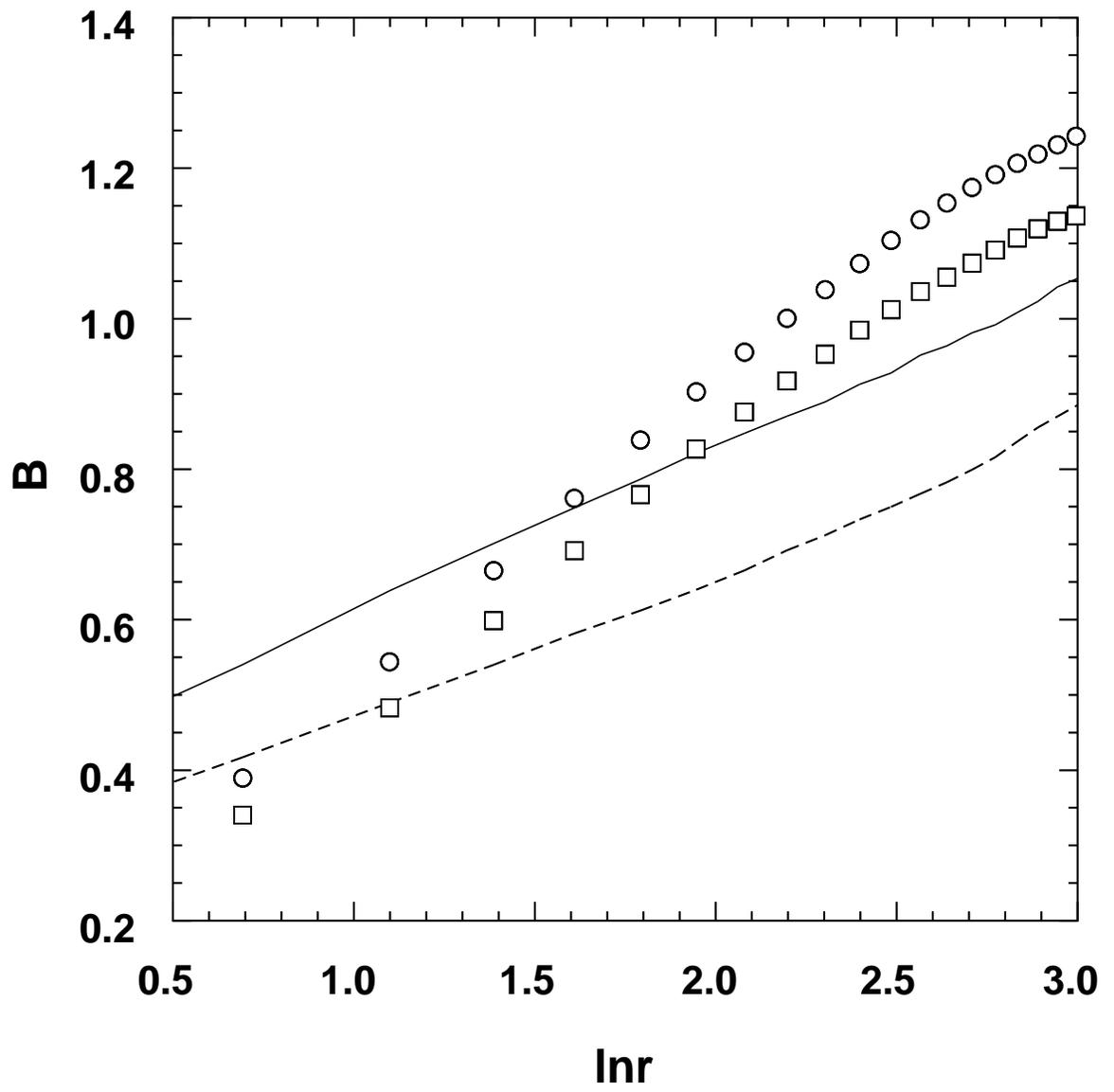
\noindent FIG. 2. $B(r)$ versus $\ln r$ for $M=60$ and $\sigma = 0.2$.
Solid line: system in equilibrium at $T=0.3$; broken line: $T=0.2$; circles:
system cooled to $T=0.01$ at $\Gamma = 1$; squares: $\Gamma = 0.5$.
\vspace{1em}

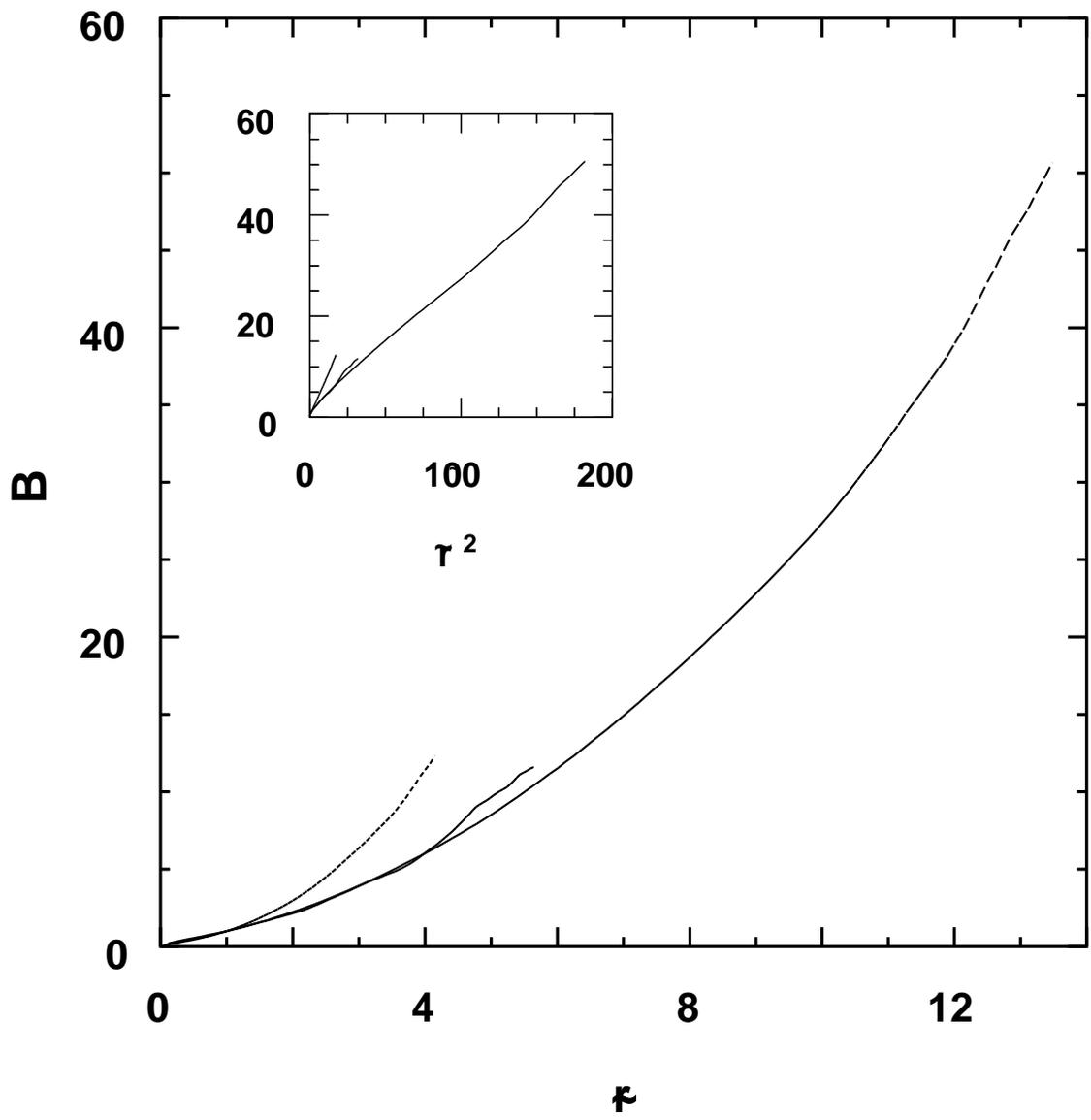
\noindent FIG. 3. Mean-square displacement $B(r)$ versus 
$\tilde{r} \equiv r/\xi$.  Dashed line: $M=120$, $\sigma=0.5$,
$\Gamma = 5 \times 10^{-4}$; solid line: $M=120$, $\sigma=0.2$,
$\Gamma = 10^{-3}$; dotted line: $M=60$, $\sigma = 0.2$,
$\Gamma = 10^{-4}$.
\vspace{1em}

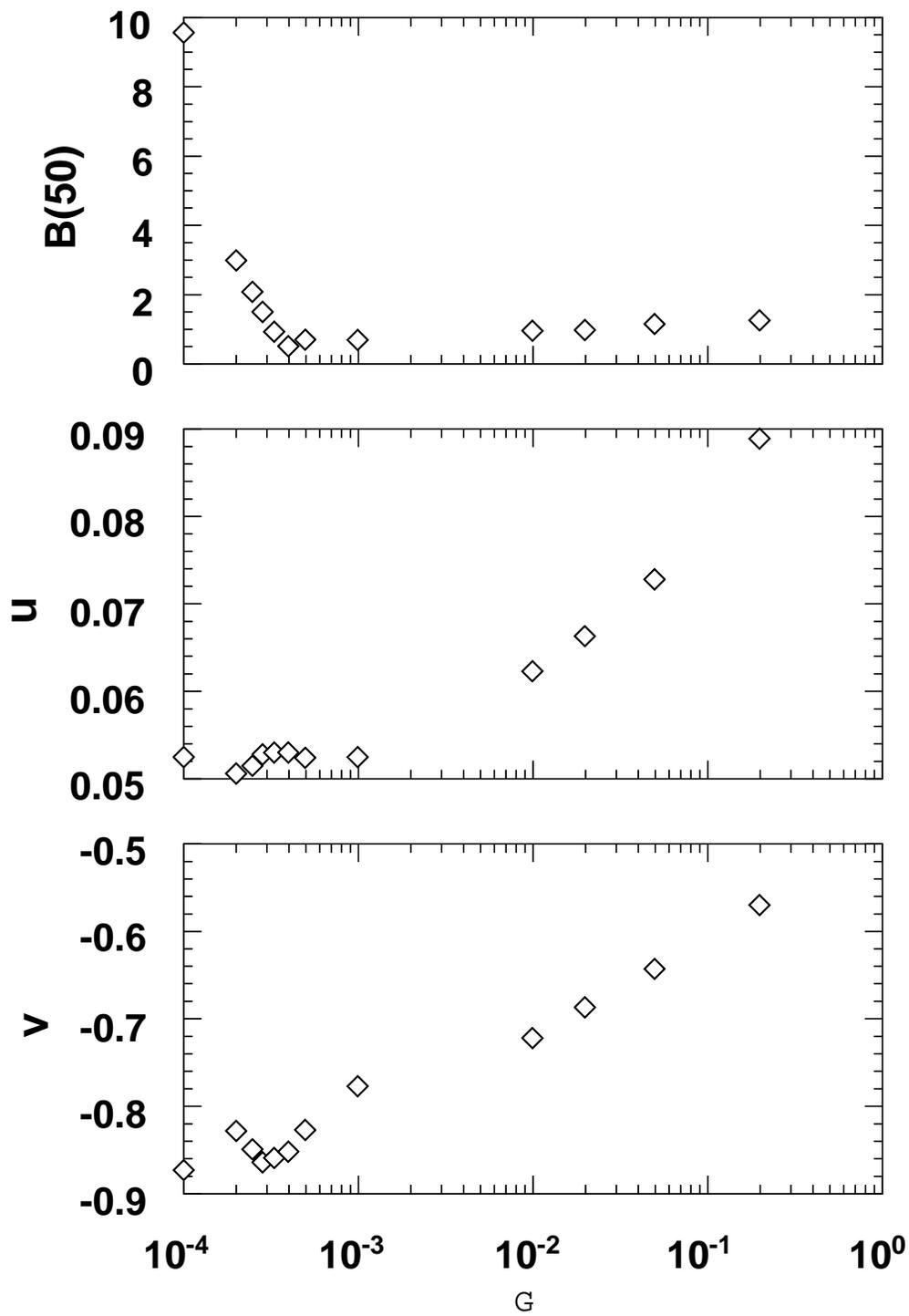
\noindent FIG. 4. Per-particle potential energy, $v$, elastic energy, $u$,
and $B(r=50)$ versus $\Gamma$ for the same parameters as in Fig. 1.
\vspace{1em}

\end{document}